\documentclass{article}

\usepackage[aux]{rerunfilecheck}
\usepackage{amsmath}
\usepackage{amssymb}
\usepackage{bm}
\usepackage{mathtools}
\usepackage[numbers,sort&compress]{natbib}
\usepackage[autostyle]{csquotes}
\usepackage{graphicx}
\usepackage[usenames,dvipsnames]{color}
\usepackage{bbm} 
\usepackage{placeins} 
\usepackage{booktabs}
\usepackage{siunitx}
\usepackage[labelfont=bf]{caption}
\usepackage[top=1in, bottom=1.25in, left=0.9in, right=0.9in]{geometry}
\usepackage{subfigure}
\usepackage{tabularx}
\usepackage{lipsum}
\usepackage{multirow}
\usepackage[hidelinks]{hyperref}
\usepackage{braket}

\title{\bf Fitting the DESI BAO Data with Dark Energy Driven by the Cohen--Kaplan--Nelson Bound}

\author{ Patrick Adolf$^1$\footnote{patrick.adolf@tu-dortmund.de}, Martin Hirsch$^2$\footnote{mahirsch@ific.uv.es}, Sara Krieg$^1$\footnote{sara.krieg@tu-dortmund.de}, Heinrich P\"as$^1$\footnote{heinrich.paes@tu-dortmund.de}, Mustafa Tabet$^1$\footnote{mustafa.tabet@tu-dortmund.de}
\smallskip
\\
{\it $^1$Fakult\"at f\"ur Physik,
Technische Universit\"at Dortmund, D-44221 Dortmund, Germany} \\
{\it $^2$Instituto de F\`isica Corpuscular (IFIC), Universidad de Valencia-CSIC,}\\
{\it E-46980 Valencia, Spain}
}

\begin{document}

\maketitle

\begin{abstract}
    Gravity constrains the range of validity of quantum field theory. As has been pointed out by Cohen, Kaplan, and Nelson (CKN), such effects lead to interdependent ultraviolet (UV) and infrared (IR) cutoffs that may stabilize the dark
    energy of the universe against quantum corrections, if the IR cutoff is set
    by the Hubble horizon. As a consequence of the cosmic expansion, this argument implies a time-dependent dark energy density. 
    In this paper we confront this idea with recent data from DESI BAO, Hubble and supernova measurements. We find that the CKN model provides a better fit to the data than the $\Lambda$CDM model and can compete with other models of time-dependent dark energy that have been studied so far.  
\end{abstract}

\section{Introduction}
\label{sec:intro}

Quantum field theory (QFT) describes physics only in the absence of strong gravitational effects. A prominent example
for the breakdown of QFT when this condition does not apply is that the maximum information stored in
a space-time region scales with the volume in QFT while it scales holographically with the horizon area for black 
holes~\cite{Bekenstein:1973ur,Bekenstein:1974ax, Bekenstein:1980jp,Bekenstein:1993dz,Hawking:1975vcx,Hawking:1976de}. 
This consideration implies that QFT overcounts fundamental degrees of freedom for large energies and large length scales and has 
been employed by Cohen, Kaplan and Nelson (CKN)~\cite{Cohen:1998zx} to derive ultraviolet (UV) and infrared (IR) cutoffs 
that restrict the range of validity of QFT. Since the entropy of a black hole provides an upper bound for the entropy for any system, 
the entropy $S_\text{QFT}$ within a box of size $L$ described by QFT is constrained by the
Bekenstein--Hawking entropy $S_\text{BH}$
\begin{equation}
    S_\text{QFT} = \Lambda^3_\text{UV} L^3\, \leq \pi L^2 M_\text{Pl}^2 = S_\text{BH}\,,
    \label{eqn:Bekenstein}
\end{equation}
where $\Lambda_\text{UV}$ represents the UV cutoff, the maximum length $L_\text{max}$
for a fixed $\Lambda_\text{UV}$
defines the IR cutoff $\Lambda_\text{IR} = L^{-1}_\text{max}$ (and vice versa),
and $M_\text{Pl}$ denotes the Planck mass. 
However, as CKN have further pointed out, an effective field theory satisfying the Bekenstein bound of equation~\eqref{eqn:Bekenstein}
still allows for the existence of numerous states with a Schwarzschild radius larger than the size of the box. 
This problem can be resolved by introducing a more stringent limit that excludes all states describing a black hole, 
the so-called ``CKN bound''~\cite{Cohen:1998zx} 
\begin{equation}
    \Lambda^4_\text{UV} \lesssim \frac{1}{L^2} M_\text{Pl}^2\,.
    \label{eqn:CKNBound}
\end{equation}
While there exist several works that study the consequences of the 
CKN bound for particle phenomenology, such as the magnetic moment of 
leptons~\cite{Cohen:2021zzr,Banks:2019arz,Bramante:2019exc,Blinov:2021fzl,Bramante:2019uub}, 
the hierarchy problem~\cite{Kephart:2022vfr} and the phenomenology of radiative neutrino masses~\cite{Adolf:2023wfw},
one of the first applications of the CKN bound that has already been discussed in the original 
CKN paper~\cite{Cohen:1998zx} is the cosmological constant problem~\cite{Weinberg:1988cp}.
After the recent result of the Dark Energy Spectroscopic Instrument (DESI)~\cite{DESI:2024mwx} in combination with other
cosmological data has provided new hints that the dark energy density of the universe may be time-dependent,
we focus here on the consequences of the CKN bound for the evolution of dark energy. 
According to QFT, quantum corrections to the dark energy density 
scale with $\sim \Lambda_{\text{UV}}^4$, where $\Lambda_{\text{UV}}$ is the UV cutoff of the corresponding theory. 
If the Planck scale is chosen as the UV cutoff of the Standard Model, this leads to a correction that is many orders of magnitudes 
larger than the observed dark energy density. One possibility to resolve this problem is to assume that QFT ceases to work at a cutoff scale 
 $\Lambda_{\text{UV}} \ll M_\text{Pl}$. Accordingly,
 CKN have suggested to adopt the Hubble horizon, i.e.~the inverse Hubble parameter as the IR cutoff of QFT, $L_\text{max}=1/H$, 
resulting in a UV cutoff $\Lambda_{\text{UV}} \sim 10^{-3} \,\text{eV}$ that corresponds to the observed dark energy density 
$\rho_{\Lambda} \sim ( 10^{-3} \,\text{eV})^4$ observed today. While this scenario may stabilize the dark energy of the universe 
against large quantum corrections, it also entails the prediction that these quantum corrections to the dark energy density are
time dependent, as the Hubble parameter is not a constant.

In the following, we review evolving dark energy due to the CKN bound, address arguments in the literature that the model
may not produce the correct equation of state for an accelerated universe, and perform a global analysis that fits the model 
to the recent DESI BAO and supernova datasets. For other recent works that discuss evolving dark energy 
explanations of the DESI data, see for example~\cite{Notari:2024rti, Gialamas:2024lyw, DES:2024fdw, VanRaamsdonk:2024sdp, Roy:2024kni,
Yang:2024kdo,Heckman:2024apk,Wang:2024dka,Wang:2024hwd,Wang:2024rjd,Wang:2024hks}. 

This paper is organized as follows: In section~\ref{sec:sec2}, we solve the
Friedmann equations for the CKN case, and provide a brief overview of alternative
dark energy models against which we compare the CKN model. 
In section~\ref{sec:sec3}, we introduce the statistical approach adopted and present the 
results of a global analysis that confronts evolving dark energy models with 
DESI BAO, Hubble and supernova data and compares them to the standard $\Lambda$CDM paradigm that describes 
cosmology with 
a dominating constant dark energy density or cosmological constant $\Lambda$ and a subdominant component of 
cold dark matter (CDM). We also discuss the expected improvements from new data anticipated for the next few years.
Finally, we summarize and discuss our findings in section~\ref{sec:dis}.

\section{Evolving Dark Energy Models and the CKN Bound}
\label{sec:sec2}

The recent analysis of measurements of the DESI baryonic acoustic oscillations (BAO) data, combined with data 
from the cosmological microwave background (CMB) and Pantheon+, Union3, or DES-SN5YR datasets probing 
supernova distances results in a  
$3.9 \sigma$ evidence for time-varying dark energy models compared to the $\Lambda$CDM paradigm~\cite{DESI:2024mwx}.
In the following, we analyze the evolution of the dark energy density in
the presence of a UV and IR cutoff satisfying the CKN bound in equation~\eqref{eqn:CKNBound}, and compare the
behavior with alternative evolving dark energy models.

Adopting an IR cutoff of the Hubble horizon size, it follows from equation~\eqref{eqn:CKNBound}
that
\begin{equation}
    \Lambda^4_\text{UV} \lesssim H^2 (z) M^2_\text{Pl}\,,
    \label{eqn:CKN}
\end{equation}
with the redshift $z$, Planck mass $M_\text{Pl}$, and the Hubble constant
$H(z)$. This in turn yields for the vacuum energy density (VED) $\rho^\text{1-loop}_\mathrm{VED}$
at 1-loop
\begin{equation}
    \rho^\text{1-loop}_\text{VED} (z) \simeq \nu\frac{\Lambda_\text{UV}^4}{16 \pi^2} \, = \nu\frac{M_\text{Pl}^2 H^2(z)}{16 \pi^2} \,.   
    \label{eqn:Corr}
\end{equation}
Here, an additional parameter $\nu$ has been introduced, such that the
VED of the original CKN bound is recovered for $\nu = 1$. In the following, we refer to $\nu = 1$ as the CKN case; otherwise, we refer to the $\nu$CKN case.

This generalization serves two purposes. First, in the derivation of equation~\eqref{eqn:CKN}, several
prefactors have been dropped such as a $1/2$ coming from the Schwarzschild radius. Further, the 
derivation of equation~\eqref{eqn:Corr} depends on the particle content of the underlying theory, 
such as possible dark matter candidates. 
Moreover, the more general equation~\eqref{eqn:Corr}
applies also for other dark energy models 
 discussed in the literature where the vacuum energy
scales proportional to $H^2(z)$ such as running vacua models~\cite{Rezaei:2019xwo, Rezaei:2021qwd, Shapiro:2000dz} or 
holographic dark energy models~\cite{Li:2004rb, Telali,Tyagi:2024cqp}, see reference~\cite{Abdalla:2022yfr} for a review. 

To arrive at equation~\eqref{eqn:Corr}, we have neglected the IR cutoff
$\Lambda_\mathrm{IR}^4\propto H(z)^4$. Given that we are only analyzing
data where $H(z)^4 \ll M_\text{Pl}^2 H(z)^2$, this simplification is justified.
Note, that we neglect the neutrino masses and that heavy particles with masses above the 
UV cutoff do not contribute~\cite{Cohen:1998zx,Cohen:2021zzr}. 

While it has been argued in the literature that a scaling proportional to $H^2(z)$ according to equation~\eqref{eqn:Corr}
leads to a wrong equation of state that does not explain the accelerated expansion of the universe~\cite{Hsu:2004ri,Li:2004rb}, this 
conclusion holds only as far as the conservation of energy is assumed to hold separately for the individual matter and 
dark energy contributions. Dropping this assumption yields a modified equation of state 
(see e.g.~\cite{Sola:2013gha,Moreno-Pulido:2022upl,Pavon:2005yx})
that leads to accelerated expansion and does in fact, as we will show, provide an excellent fit to data in the range of redshifts up to $z \sim {\cal O}(1)$.

Moreover, it
has been pointed out in reference~\cite{Hsu:2004ri} that a $H^2(z)$ scaling may conflict with processes at large $z$ such as CMB data or large structure formation. However, this behavior may be alleviated by a matter component starting to become dominant around 
$z \gtrsim 0.4$ and a small prefactor $\nu \lesssim 1$ in the $\nu$CKN model. In fact, as the prefactor depends in general on the concrete particle physics model, it is unknown how the parameter $\nu$ evolves with time or redshift.

To study the consequences of the CKN bound for the evolution of the universe, 
we introduce the correction term from equation~\eqref{eqn:Corr} semi-classically into the energy--momentum tensor $T^{\mu\nu}$~\cite{Weinberg:1988cp}: 
\begin{align}
    T^{\mu\nu}_\text{tot} &= T^{\mu\nu}_\text{classical} + \braket{T^{\mu\nu}},
\end{align}
with
\begin{align}
    \braket{T^{\mu\nu}}
    &= \rho^\text{1-loop}_\text{VED} g^{\mu\nu}.
\end{align}
Here, $T^{\mu\nu}_\text{classical}$ contains the contributions from matter, radiation and 
the classical cosmological constant $\Lambda_0$
\begin{align}
  T^{\mu\nu}_\text{classical}
  &= T^{\mu\nu}_\text{matter} + T^{\mu\nu}_\text{radiation} + \Lambda_0 g^{\mu\nu}.
\end{align}
It is important to note that matter and dark
energy are no longer conserved separately but can transform into each other. Such effects of matter non-conservation 
have been discussed for example in references~\cite{Sola:2013gha,Pavon:2005yx}.
Otherwise, the Bianchi identity $\nabla_\mu G^{\mu\nu} = 0$, where $G$ denotes the Einstein tensor, together with the 
conservation of the classical energy--momentum tensor, would imply that: 
\begin{equation}
    \nabla_\mu \braket{T^{\mu\nu}} = 0 \Rightarrow \dot{\rho}^\text{1-loop}_\text{VED} = 0\,,
\end{equation}
which is in contradiction with the time dependence of the VED, cf.
reference~\cite{SolaPeracaula:2022hpd} for further discussion. Combining the equation of the $\mu\nu=00$ and $\mu\nu=ij$ components of the Einstein equation leads to the usual Friedmann equation
\begin{equation}\label{eqn:hubble-sq}
    H^2(t) = \frac{8 \pi G}{3} \left(\rho_\text{M}(t) + \rho_\Lambda (t)\right)\,,
\end{equation}
with the matter density $\rho_\text{M}(t)$, the dark energy density
$\rho_\Lambda(t) = \rho^\text{1-loop}_\text{VED}(t) + \Lambda_0$ and Newton's constant $G$. Moreover, we neglect radiation
as we are only interested in the matter-dominated era, and
consider a spatially flat universe, i.e.~$\Omega_\text{k} = 0$. The conservation of
the zeroth component of the energy--momentum tensor, i.e.~$\nabla_\mu {T^\mu}_0 = 0$, yields
\begin{equation}\label{eqn:emt-conservation}
    \dot{\rho}_{\Lambda} (t) + \dot{\rho}_\text{M}(t) = -3 H(t) \rho_\text{M}(t)\,,
\end{equation}
where we already made use of the equation of state for the matter component, $p_m = \omega_m \rho_m$ with $\omega_m = 0$.
Solving equations~\eqref{eqn:hubble-sq} and~\eqref{eqn:emt-conservation} we obtain
\begin{equation}
    H^2(z) = H_0^2 \left(\Omega_\text{M}(z) + \Omega_\Lambda(z)\right)\,,   
\end{equation}
where $H_0 \equiv H(z = 0)$ denotes today's value of the Hubble constant and
\begin{align}
    \Omega_\text{M} (z) &= \Omega_\text{M}^0 \left(1+z\right)^{3 - \frac{\nu}{2 \pi}}\,,\\
    \Omega_\Lambda (z) &= \Omega_\Lambda^0 + \Omega_\text{M}^0 \frac{\nu}{6 \pi - \nu}\left[\left(1+z\right)^{3 - \frac{\nu}{2 \pi}} - 1\right]\,,
\end{align}
where $\Omega_\text{M}^0$ and $\Omega_\Lambda^0$ denote the matter and dark
energy density today normalized to $\rho_\text{crit,0} = 3 H_0^2/(8 \pi G)$,
respectively. 
For $\nu > 0$ one finds a quintessence-like behavior for the dark energy~\cite{Zlatev:1998tr}.

In the following, we briefly discuss several alternative
dark energy models that we are going to
compare with the CKN and $\nu$CKN cases in section~\ref{sec:sec3}.
The Friedmann equation for a spatially flat universe in the matter dominated
epoch for the most common models can usually be written as follows~\cite{DESI:2024kob}:
\begin{equation}
    \frac{H^2(z)}{H_0^2} = \Omega_\text{M}^0 \left(1+z\right)^{3} + f_\text{DE}(z) \,,
    \label{eqn:Gen_Hubble}
\end{equation}
where the matter density parameter $\Omega_\text{M}^0$ indicates the proportion
of matter relative to the critical density of the universe today, $\rho_\text{crit,0}$, and
$f_\text{DE}(z)$ describes the dark energy evolution in the universe.
One can define a relationship between the density $\rho$ and pressure $p$, expressed through a parameter $\omega$
as $p = \omega \rho$. When $\omega$ is a constant, this equation is referred to as equation of state.  
In the general case $\omega$ can be written as:
\begin{equation}\label{eqn:not-so-eos}
    \omega(z) = -1 + \frac{1}{3}\frac{\mathrm{d}\,\mathrm{ln}f_\text{DE}(z)}{\mathrm{d}\,\mathrm{ln} \left(1+z\right)}\,,
\end{equation}
where $\omega < -1/3$ leads to an accelerated expansion of the universe corresponding to dark energy.
Several cosmological models result from different values and parametrizations of $\omega$.
In particular, we are interested in the following three models that we compare against the CKN
cases in the subsequent section.
\paragraph*{$\bm{\Lambda}$CDM Model} In the standard cosmological model, a cosmological constant $\Lambda$ 
is introduced into the Einstein field equations, which in turn leads to a constant dark energy. This 
corresponds to the choice of $\omega = -1$, i.e.~$p = -\rho$, which yields
\begin{equation}
    f_{\text{DE}}^{\Lambda\text{CDM}}(z) = \Omega_{\text{DE},0}^{\Lambda\text{CDM}} \,,
\end{equation}
where $\Omega_{\text{DE},0}^{\Lambda\text{CDM}}$ denotes the present-day dark energy density in the $\Lambda$CDM 
model. 
\paragraph*{$\bm{\omega}$CDM Model} In addition to matter and radiation, there may be
for example other potential sources 
contributing to the energy--momentum tensor, which can only be explained by $\omega\neq -1$
like e.g.~scalar fields in inflation models~\cite{Barrow:2019gup}. This results
in a dark energy that is no longer constant in time. Therefore, $\omega$ becomes a free parameter of the model and 
one can derive
\begin{equation}
    f_{\text{DE}}^{\omega\text{CDM}}(z)
        = \Omega_{\text{DE},0}^{\omega\text{CDM}} \left(1+z\right)^{3\left(1+\omega\right)}\,,
\end{equation}
where $\Omega_{\text{DE},0}^{\omega\text{CDM}}$ is the present-day dark energy density in the $\omega$CDM model. If 
experiments show that $\omega \neq -1$, this would indicate a time-varying dark energy. 
\paragraph*{$\bm{\omega_0 \omega_a}$CDM Model} To incorporate the dynamics of dark
energy into a model, the parametrization $\omega (a) = \omega_0 + \omega_a (1 -a)$
has been derived~\cite{Chevallier:2000qy,Linder:2002et}, where $a = (1+z)^{-1}$
is the scale factor. This leads to
\begin{equation}
    f_{\text{DE}}^{\omega_0\omega_a\text{CDM}}(z)
        = \Omega_{\text{DE},0}^{\omega_0\omega_a\text{CDM}} (1+z)^{3\left(1+\omega_0+\omega_a\right)} e^{-3\omega_a\left(1-1/(1+z)\right)}\,,
\end{equation}
with the present-day dark energy density $\Omega_{\text{DE},0}^{\omega_0\omega_a\text{CDM}}$ in the
$\omega_0\omega_a\text{CDM}$ model. It is evident that for $\omega_a = 0$ one recovers the $\omega$CDM model.\\
\\
In the following, we assume for all models $\Omega_\text{M}^0 + \Omega_\text{DE,0} = 1.$

\section{Comparison with Experimental Data}
\label{sec:sec3}
In this section, we perform a global analysis to determine the agreement of
the CKN bound with current experimental data in particular in light of the recent
DESI measurements~\cite{DESI:2024mwx}.
Firstly, in section~\ref{subsec:sec31} we list and briefly discuss the
datasets we use while in section~\ref{subsec:sec32} 
the statistical approach is explained. Finally, our results are 
presented in section~\ref{subsec:sec33}.
\subsection{Experimental Input}
\label{subsec:sec31}
In our numerical analysis, the dataset from DESI BAO~\cite{DESI:2024mwx} is used in combination with
model-independent Hubble measurements~\cite{Favale:2024lgp}.
Additionally, two supernova distance datasets are used: Pantheon+~\cite{Brout:2022vxf},
and the DES-SN5YR~\cite{DES:2024tys} dataset which we denote as DESY5 in the following.
\paragraph*{DESI BAO} Baryonic acoustic oscillations are fluctuations in the density of baryonic matter,
caused by acoustic density waves in the primordial plasma of the early universe.
The corresponding observables are extracted from galaxy, quasar and Lyman-$\alpha$ forest tracers, and can be found in
reference~\cite{DESI:2024mwx}. There, they provide measurements and correlations for the comoving distance over the 
drag epoch $D_\mathrm{M}/r_\mathrm{d}$ and the distance variable $D_\mathrm{H}/r_\mathrm{d}$. 
Here, $r_\text{d}$ denotes the drag epoch, which is the
distance sound can travel in the time from the Big Bang till the decoupling of
the baryons. 
In cases with low signal-to-noise ratio instead the angle-average quantity $D_\mathrm{V} / r_\mathrm{d}$
is provided. Note, that with DESI BAO data alone only the combination $r_\text{d} H_0$ and the parameter $\Omega_\text{M}^0$ can be
constrained. Here, we keep the drag epoch $r_\mathrm{d}$ a free parameter in the fit. 
The dataset consists of seven bins with negligible correlations, cf. reference~\cite{DESI:2024mwx},
between the overlapping bins.

\paragraph*{Supernova} Type Ia supernovae are considered as standard candles for determining cosmological
distances. When distances are measured, Hubble's parameter can be inferred using the formula for the physical distance. 
Here, we consider the datasets from Pantheon+ and DESY5.
Unfortunately, the Union3 data is not publicly available.

For DESY5 the distance modulus and corresponding covariance matrices are taken from reference~\cite{DES:2024tys}.
Note, that the unknown absolute magnitude $M$ always appears with the Hubble constant $H_0$ such that the supernovae
dataset alone cannot be used to determine $H_0$.
Thus, analogous to reference~\cite{DES:2024tys}, we redefine the absolute magnitude $M$ and Hubble parameter $H_0$
into one single parameter $\tilde{M} = M + 5\log_{10}(c/H_0)$ over which we marginalize analytically. Here, $c$ denotes
the speed of light.

The Pantheon+ data and corresponding covariance matrices are taken from reference~\cite{Brout:2022vxf}.
Here, again, we analytically marginalize over $\tilde{M}$ and analogous to reference~\cite{DESI:2024mwx}, 
we consider supernovae with redshifts $z>0.01$. 

\paragraph*{Model-Independent Hubble Parameter Measurements} An additional method,
independent of specific cosmological models, exists for measuring the 
Hubble parameter. The data for these measurements is taken 
from reference~\cite{Favale:2024lgp} and is based on cosmic chronometers. Unfortunately, only the covariance matrix of Moresco et al.~is provided~\cite{Moresco:2020fbm}
which, however, also contains the biggest source of correlation. Therefore, we neglect the correlation between the other measurements.\\
\\
Note, that for this initial analysis, we neglect CMB data from temperature, polarization, 
and lensing measurements.
In our approach, we fit all three experiments simultaneously, incorporating one
supernova dataset at a time. By including the model-independent Hubble
measurements, we also lift the afore-mentioned degeneracies in the parameters.

\subsection{Statistical Procedure}
\label{subsec:sec32}
To quantify the agreement between the different models, in particular the CKN cases and the experimental datasets listed in section~\ref{subsec:sec31}, we 
perform a $\chi^2$ test. In order to determine the best-fit point, we minimize the function 
\begin{equation}\label{eqn:chi2}
    \chi^2 = \left(\vec{\mathcal{O}}_\text{th} \left(\xi_i\right) - \vec{\mathcal{O}}_\text{exp} \right)^T C^{-1} \left(\vec{\mathcal{O}}_\text{th} \left(\xi_i\right) - \vec{\mathcal{O}}_\text{exp} \right)\,,
\end{equation}
where the vector $\vec{\mathcal{O}}_\text{th}$ denotes the theory predictions as a function of the model parameters 
$\xi_i$, while the corresponding measurements with the covariance matrix $C$ are denoted by the vector $\vec{\mathcal{O}}_\text{exp}$.
Note, that equation~\eqref{eqn:chi2} in its compact form does not yet account
for the analytical marginalization we perform for the DESY5 and Pantheon+ likelihood,
also see references~\cite{DES:2024tys, Goliath:2001af}.
Despite the Bayesian nature of the underlying datasets, a $\chi^2$ statistics is a reasonable
approximation as the provided uncertainties are Gaussian and the theoretical models are linear
to a good approximation. Thus, we estimate the uncertainties and covariance matrix using the 
Hessian matrix at the minimum of the $\chi^2$ distribution. We also provide the frequentist 
confidence levels (CL) by calculating the corresponding $\chi^2$ contours
which are determined for a given parameter plane of interest, by minimizing
the $\chi^2$ function for every grid point with respect to the remaining
parameters. The analysis has been performed in \texttt{Mathematica} and
the global minimizations are done using differential evolution.

We also compare CKN with other cosmological models, namely $\Lambda$CDM, $\omega$CDM and $\omega_0 \omega_a$CDM 
introduced in section~\ref{sec:sec2}, by calculating the difference $\Delta \chi^2 = \chi^{2, (\nu)\mathrm{CKN}}_\text{min} - \chi^{2, \mathrm{alt. model}}_\text{min}$. 
For this comparison and assuming that Wilk's theorem holds, $\Delta \chi^2$ follows a $\chi^2$ distribution 
corresponding to the difference in the number of model parameters, respectively. 
This allows us to convert the $\Delta\chi^2$ values into the corresponding significances in terms
of the one dimensional normal distribution. Note, however, that Wilk's theorem only holds for comparisons
between nested models. Therefore, we can only convert the $\Delta\chi^2$ values between the $\nu$CKN case and the $\Lambda$CDM model into a significance this way. In order to also quantify the agreement between the non-nested models, we compare the models using the 
differences in the Akaike information criterion (AIC) that quantifies 
the quality of the models fitting the data and also penalizes an increasing number of model parameters~\cite{1100705} with 
\begin{equation}
    \text{AIC} = \chi^2_\text{min} + 2k\,,
\end{equation}
where $k$ is the number of model parameters. 

\subsection{Results}
\label{subsec:sec33}
\begin{figure}
    \centering
        \includegraphics[scale=1]{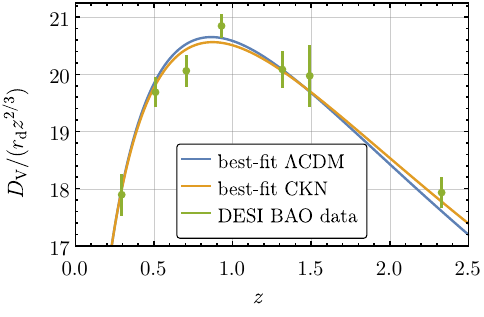}
        \includegraphics[scale=1]{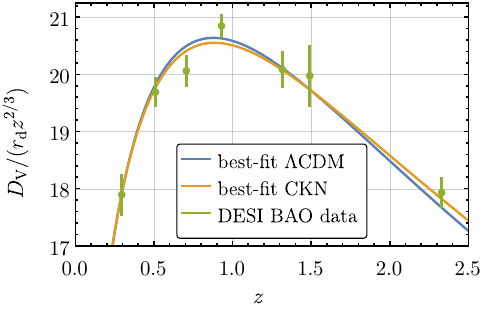}
    \caption{Shown is the angle-averaged distance quantity $D_\text{V}/(r_\text{d} z^{2/3})$ at our best-fit point together with the 
    corresponding DESI measurements~\cite{DESI:2024mwx} for both, the $\Lambda$CDM model and the CKN model. For the best-fit points the DESY5 dataset (left) and Pantheon+ dataset (right) are used. Since the results of the CKN and $\nu$CKN model are too similar to see any difference, 
    we only show the CKN model.}
    \label{fig:DESI}
\end{figure}

In table~\ref{tab:CKN_with_data} we show the best-fit points of the 
CKN model to the DESI BAO+Hubble datasets, each combined with either DESY5 or Pantheon+. The individual 
contribution of each experiment is given in table~\ref{tab:Pulls}.
\begin{table}
    \caption{Results of the best-fit points for the CKN and $\nu$CKN case to the datasets of DESI BAO and Hubble, once
    combined with DESY5 and once with Pantheon+ data. Shown are the results for Hubble-today $H_0$, the matter density 
    parameter $\Omega_\text{M}^0$, the drag epoch $r_\text{d}$, the parameter $\nu$ and the minimal $\chi^2_\text{min}$ over the degrees of freedom (DOF).}
    \begin{center}
    \begin{tabular}{l c c c c c} 
    \toprule
    {\bf Model}/Datasets & $H_0/(\text{km/s/Mpc})$ & $\Omega_\text{M}^0$ & $r_\text{d}/\text{Mpc}$ & $\nu$ & $\chi^2_\text{min}/\text{DOF}$\\
    \midrule
    \bf{CKN}\\
    + DESY5 & $\num{68.69}\pm\num{2.39}$ & $\num{0.354}\pm\num{0.012}$ & $\num{144.54}\pm\num{4.89}$ & -- & 1677/1871 \\
    + Pantheon+ & $\num{69.24}\pm\num{2.41}$ & $\num{0.344}\pm\num{0.012}$ & $\num{144.38}\pm\num{4.89}$&-- & $\num{1440}/1632$\\
    \midrule
    \bf{$\bm{\nu}$CKN}\\
    + DESY5 & $\num{68.66}\pm\num{2.38}$ & $\num{0.354}\pm\num{0.021}$ & $\num{144.60}\pm\num{4.87}$ & $\num{1.00}\pm\num{0.46}$ & 1677/1870 \\
    + Pantheon+ & $\num{67.94}\pm\num{2.51}$ & $\num{0.329}\pm\num{0.021}$ & $\num{147.76}\pm\num{5.28}$& $\num{0.60}\pm\num{0.49}$ & $\num{1440}/1631$\\
    \bottomrule
    \end{tabular}
    \label{tab:CKN_with_data}
    \end{center}
\end{table}
\begin{table}
    \caption{Contribution of the different experiments to the total $\chi^2_\text{min}$ from table~\ref{tab:CKN_with_data} for
    the combination with DESY5 and Pantheon+.}
    \begin{center}
    \begin{tabular}{c c c c c c c} 
    \toprule
    \multirow{2}{*}{Models} & \multicolumn{3}{c}{DESY5} & \multicolumn{3}{c}{Pantheon+}\\
    \cmidrule(l){2-4} 
    \cmidrule(l){5-7}
    & $ \chi^{2, \text{BAO}}_\text{min}$ & $\chi^{2, \text{DESY5}}_\text{min}$ & $\chi^{2, \text{Hubble}}_\text{min}$ & $ \chi^{2, \text{BAO}}_\text{min}$ & $\chi^{2, \text{Pantheon+}}_\text{min}$ & $\chi^{2, \text{Hubble}}_\text{min}$\\
    \midrule
    CKN & 14.48 & 1649 & 12.86 & 14.37 & 1413 & 12.67\\
    $\nu$CKN & 14.48 & 1649 & 12.85 & 13.51 & 1413 & 13.08\\
    \bottomrule
    \end{tabular}
    \label{tab:Pulls}
    \end{center}
\end{table}
It is already evident from the $\chi^2_\text{min}/\text{DOF}$ values that the CKN as well as the $\nu$CKN model are well compatible 
with the data. This is substantiated by the best-fit which for the CKN as well as the $\nu$CKN case is given by
$\chi^2_\text{min}/\text{DOF} \approx 0.90$ for the DESY5 dataset and $\chi^2_\text{min}/\text{DOF} \approx 0.88$ 
for the Pantheon+ dataset.
Additionally, we show the angle-averaged distance quantity $D_\text{V}/(r_\text{d} z^{2/3})$ at our best-fit point together with the corresponding DESI measurements in figure~\ref{fig:DESI}.

The quantitative comparison of the CKN and the $\nu$CKN model with the $\Lambda$CDM, $\omega$CDM and $\omega_0 \omega_a$CDM model is 
presented in table~\ref{tab:CKN_with_models}. There, we provide the difference $\Delta \chi^2$ of the best-fit points and the difference in AIC. 
The other models are fitted using the same approach as for the 
CKN and $\nu$CKN case. The resulting best-fit points are presented in table~\ref{tab:LCDM_with_data} and the procedure was compared with 
similar analyses in the literature where we find excellent agreement.
\begin{table}
    \caption{Comparison between CKN and $\nu$CKN and the alternative cosmological models. 
    Shown is the difference 
    $\Delta \chi^2 = \chi^{2, (\nu)\mathrm{CKN}}_\text{min} - \chi^{2, \mathrm{alt. model}}_\text{min}$ for 
    both datasets and the difference $\Delta$AIC. The negative 
    values denote a preference of CKN and $\nu$CKN over the alternative models, respectively. See text for a comparison of the $\nu$CKN and $\Lambda$CDM model in terms of significances.}
    \begin{center}
    \begin{tabular}{c c c c c c} 
    \toprule
    Models & $\Delta \chi^2_\text{DESY5}$  & $\Delta$AIC & & $\Delta \chi^2_\text{Pantheon+}$  & $\Delta$AIC\\
    \midrule
    \textbf{CKN with}\\
    $\Lambda$CDM           & $\num{-4.6}$  & $\num{-4.6}$  & & $\num{-1.1}$  & $\num{-1.1}$\\
    $\omega$CDM            & $\num{2.3}$   & $\num{0.3}$   & & $\num{1.5}$   & $\num{-0.5}$\\
    $\omega_0 \omega_a$CDM & $\num{5.3}$   & $\num{1.3}$   & & $\num{2.1}$   & $\num{-1.9}$\\
    \midrule
    \textbf{$\bm{\nu}$CKN with}\\
    $\Lambda$CDM           & $\num{-4.6}$ & $\num{-2.6}$ & & $\num{-1.2}$ & $\num{0.8}$\\
    $\omega$CDM            & $\num{2.3}$  & $\num{2.3}$  & & $\num{1.4}$  & $\num{1.4}$\\
    $\omega_0 \omega_a$CDM & $\num{5.3}$  & $\num{3.3}$  & & $\num{2.0}$  & $\num{0.0}$\\
    \bottomrule
    \end{tabular}
    \label{tab:CKN_with_models}
    \end{center}
\end{table}
\begin{table}
    \caption{Results of the best-fit points from the $\Lambda$CDM, $\omega$CDM and $\omega_0 \omega_a$CDM model to the datasets of DESI BAO and Hubble, once
    with DESY5 and once with Pantheon+ data. Shown are the results for Hubble-today $H_0$, the matter density 
    parameter $\Omega_\text{M}^0$, the drag epoch $r_\text{d}$, the parameters $\omega_0$ and $\omega_a$, and the minimal $\chi^2_\text{min}$ over the degrees of freedom (DOF).}
    \begin{center}
    \begin{tabular}{l c c c c c c} 
    \toprule
    {\bf Model} & $H_0$ in & $\Omega_\text{M}^0$ & $r_\text{d}$ in & $\omega$ or $\omega_0$ & $\omega_a$ & $\chi^2_\text{min}/\text{DOF}$\\
    /Datasets & $\text{km/s/Mpc}$ & & Mpc\\
    \midrule
    \textbf{$\bm{\Lambda}$CDM}\\
    + DESY5 & $\num{69.15}\pm\num{2.41}$ & $\num{0.320}\pm\num{0.011}$ & $\num{144.55}\pm\num{4.90}$ & -- & -- & 1681/1871 \\
    + Pantheon+ & $\num{69.79}\pm\num{2.44}$ & $\num{0.308}\pm\num{0.011}$ & $\num{144.41}\pm\num{4.89}$ & -- & -- & $\num{1441}/1632$\\
    \midrule
    \textbf{$\bm{\omega}$CDM}\\
    + DESY5& $\num{68.71}\pm\num{2.39}$ & $\num{0.296}\pm\num{0.014}$ & $\num{144.27}\pm\num{4.89}$ & $\num{-0.874}\pm\num{0.046}$ & -- & 1674/1870 \\
    + Pantheon+ &$\num{69.43}\pm\num{2.43}$ & $\num{0.296}\pm\num{0.014}$ & $\num{144.26}\pm\num{4.88}$ & $\num{-0.921}\pm\num{0.048}$ & -- & $\num{1438}/1631$\\
    \midrule
    \textbf{$\bm{\omega_0 \omega_a}$CDM}\\
    + DESY5 & $\num{66.93}\pm\num{2.40}$ & $\num{0.334}\pm\num{0.016}$ & $\num{147.57}\pm\num{5.17}$ & $\num{-0.716}\pm\num{0.093}$ & $\num{-1.36}\pm\num{0.65}$ & 1671/1869 \\
    + Pantheon+ & $\num{69.65}\pm\num{2.43}$ & $\num{0.312}\pm\num{0.018}$ & $\num{143.79}\pm\num{4.86}$ & $\num{-0.873}\pm\num{0.076}$ & $\num{-0.51}\pm\num{0.58}$ & $\num{1438}/1630$\\
    \bottomrule
    \end{tabular}
    \label{tab:LCDM_with_data}
    \end{center}
\end{table}
Already from the $\Delta \chi^2$ value, it is clear that both the CKN and the $\nu$CKN case provide 
a better explanation of the data compared to the $\Lambda$CDM model, for both datasets. However, for the 
Pantheon+ dataset the difference turns out to be non-significant, and in the $\nu$CKN case the corresponding $\Delta \chi^2$ value translates into a significance of $-1.1\sigma$. 
For the DESY5 dataset, both the $\Delta \chi^2$ and AIC indicate that the $\omega$CDM and 
$\omega_0 \omega_a$CDM model perform slightly better than our models. Nevertheless, the CKN and $\nu$CKN perform better than the $\Lambda$CDM model with the $\nu$CKN model being preferred by $2.1\sigma$. 
For the Pantheon+ dataset, the CKN model actually turns out to describe the data slightly better than 
all other models according to the AIC with the largest difference between the CKN case and the $\omega_0 \omega_a$CDM model.

\begin{figure}
    \centering
        \includegraphics[scale=1]{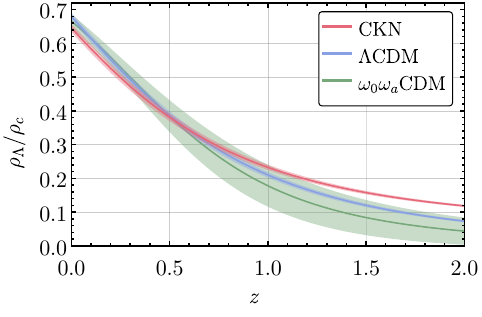}
        \includegraphics[scale=1]{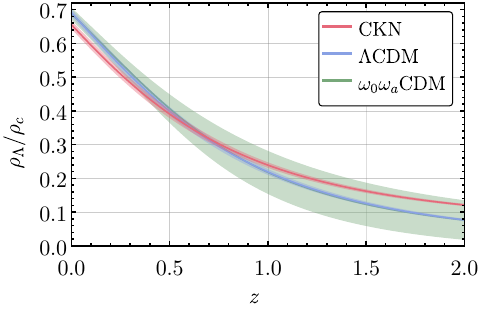}
    \caption{Shown is the evolution of the dark energy density $\rho_\Lambda$ with the $1\sigma$ band in the CKN case in comparison with the $\Lambda$CDM, 
    and $\omega_0 \omega_a$CDM model for 
    the datasets DESI BAO+Hubble+DESY5 (left) and DESI BAO+Hubble+Pantheon+ (right) normalized to the critical density $\rho_\text{c}(z)$.}
    \label{fig:CKN_other_models}
\end{figure}

\begin{figure}
    \centering
        \includegraphics[scale=1]{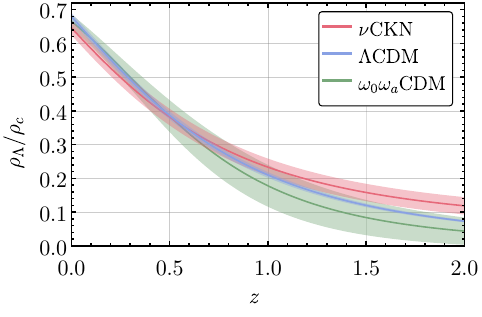}
        \includegraphics[scale=1]{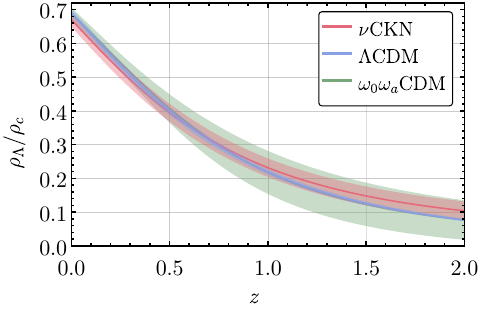}
    \caption{As figure~\ref{fig:CKN_other_models}, for the 
$\nu$CKN case.}
    \label{fig:nuCKN_other_models}
\end{figure}

We also compare the evolution of the dark energy density of the alternative models at the best-fit points with the 1$\sigma$ band in 
figure~\ref{fig:CKN_other_models} with the CKN case
and in figure~\ref{fig:nuCKN_other_models} with the $\nu$CKN case. 
It can be observed that the qualitative evolution of the dark energy density is similar across all models, 
and the corresponding dark energy densities largely overlap within their $1\sigma$ bands. 

\begin{figure}
        \includegraphics[scale=1]{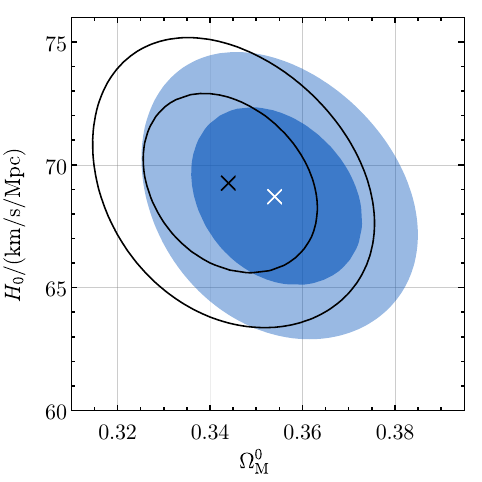} 
        \includegraphics[scale=1]{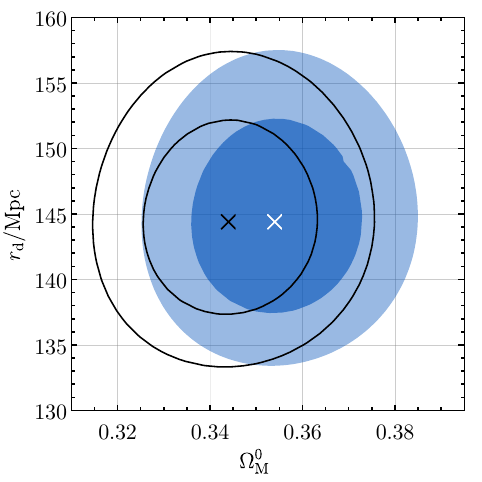}
        \includegraphics[scale=1]{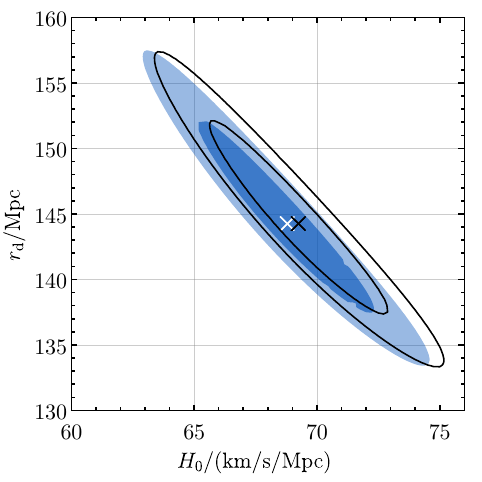} \hspace*{10em}
        \raisebox{10em}{\includegraphics[scale=1]{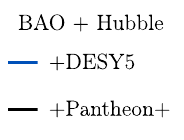}}
    \caption{Correlations of $H_0$--$\Omega_\text{M}^0$ (top left), $H_0$--$r_\text{d}$ (top right), $r_\text{d}$--$\Omega_\text{M}^0$ (bottom left) in the 
    CKN model for the DESI BAO+Hubble+DESY5 (blue area) and DESI BAO+Hubble+Pantheon+ (black lines) dataset at the 95\% and 68\% CL.}
    \label{fig:Corr_CKN}
\end{figure}

\begin{figure}
    \includegraphics[scale=1]{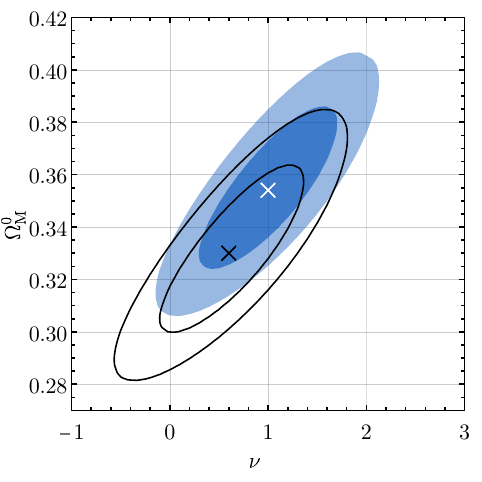} 
    \includegraphics[scale=1]{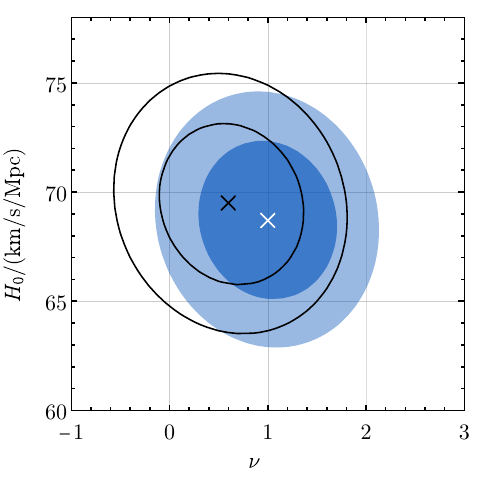}
    \includegraphics[scale=1]{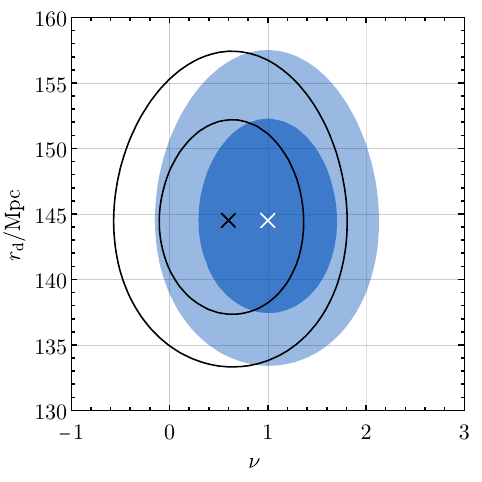} \hspace*{10em}
    \raisebox{10em}{\includegraphics[scale=1]{plots/legend.pdf}}
\caption{Correlations of $\nu$--$\Omega_\text{M}^0$ (top left), $\nu$--$H_0$ (top right), $\nu$--$r_\text{d}$ (bottom left) in the 
$\nu$CKN model for the DESI BAO+Hubble+DESY5 (blue area) and DESI BAO+Hubble+Pantheon+ (black lines) datasets at the 95\% and 68\% CL.}
\label{fig:nu_corr_CKN}
\end{figure}

Finally, the correlation at 95\% and 68\% CL for both dataset combinations for $\Omega_M^0$--$H_0$, 
$H_0$--$r_\text{d}$, $\Omega_M^0$--$r_\text{d}$ are provided in figure~\ref{fig:Corr_CKN} for the CKN.
Moreover, the correlations for $\nu$--$\Omega_M^0$, $\nu$--$H_0$, $\nu$--$r_\text{d}$ are shown in
figure~\ref{fig:nu_corr_CKN} for the $\nu$CKN case.

\subsubsection*{Future Projection}
In the near future,
DESI will continue collecting data for another three years, and other
experiments such as Euclid~\cite{EUCLID:2011zbd,Euclid:2024yrr}, which started
measuring last year, and the Large Synoptic Survey Telescope (LSST) at the Vera
Rubin Observatory will begin gathering data~\cite{LSST:2008ijt}. This will
result in significantly better statistics, smaller uncertainties, and thus allow for a 
more significant discrimination between the various cosmological models. Euclid is
expecting an improvement on the uncertainties of cosmological model parameters
by up to a factor of $10$~\cite{Euclid:2024yrr}. 

We thus perform a rough estimation of the $\chi^2$ difference between the
different models that can be anticipated with future data. We assume that the central values of
the experimental data stay the same, but with reduced uncertainties. In the
case of the DESI BAO measurements, we reduce the uncertainties of the year-1
data release by $\sqrt{5}$ in order to account for the planned 5 years duration.
In the case of the running Euclid experiment, we only consider the improvement of the uncertainties on the distance
luminosity measurements. In order to account for the expected statistics, we
rescale the uncertainties of the DESY5 and Pantheon+ datasets with a global
factor. This factor is extracted from the projected reduction of the
uncertainties on the model parameters of the $\omega$CDM
model~\cite{Euclid:2024yrr}. In order to be conservative, we take the lowest
improvement, which leads to a reduction by a factor 4. In table~\ref{tab:fp} we show
the $\chi^2$ differences between the ($\nu$)CKN and the alternative models for
the DESI projection, called DESI-5Y, the projection of the expected improvement of Euclid on the uncertainties, denoted Euclid-Unc, and the combined
projections for both, the DESY5 and Pantheon+ datasets.
As table~\ref{tab:fp} shows, if the central values of the current data remain
unchanged with future, improved measurements, the $\omega$CDM and
$\omega_0\omega_a$CDM models would be preferred.  However, we need to
stress that not only error bars but also central values of the data
points, used in this analysis, will certainly change with time and
these changes might favour or disfavour any of the models shown. Table~\ref{tab:fp},
therefore, can serve only as a demonstration of the power of future
measurements to distinguish cosmological models, but should not be
interpreted as a prediction which of the models considered in this
paper is the correct one.
In table~\ref{tab:fp2} we also show
the corresponding significances. While DESI-5Y does not show a significant difference between the models, the 
combination of DESI-5Y and the Euclid-Unc projection looks promising in distinguishing between $\Lambda$CDM and 
all time-varying dark energy models considered in this work. 
\begin{table}
    \caption{Shown is the future projection of the results of the DESI experiment after five years of measurements (DESI-5Y) and the expected improvement of the Euclid experiment on the uncertainties of the distance luminosity measurements (Euclid-Unc). We compare the $\chi^2_\text{min}$ difference between the CKN and $\nu$CKN model to the other models studied in this work for both datasets: DESY5 and Pantheon+.}
    \begin{center}
    \begin{tabular}{c c c c c c c} 
    \toprule
    \multirow{2}{*}{\bf Models} & \multicolumn{2}{c}{DESI-5Y} & \multicolumn{2}{c}{Euclid-Unc}& \multicolumn{2}{c}{DESI-5Y + Euclid-Unc}\\
    \cmidrule(l){2-3} 
    \cmidrule(l){4-5}
    \cmidrule(l){6-7}
    & $\Delta \chi^{2}_\text{DESY5}$ & $\Delta \chi^{2}_\text{Pantheon+}$ & $\Delta \chi^{2}_\text{DESY5}$  & $\Delta \chi^{2}_\text{Pantheon+}$ & $\Delta \chi^{2}_\text{DESY5}$  & $\Delta \chi^{2}_\text{Pantheon+}$\\
    \midrule
    \textbf{CKN with}\\
    $\Lambda$CDM             & $\num{-1.8}$     & $\num{3.9}$     & $\num{-16.5}$  &   $\num{-9.4}$  & $\num{-42.4}$   & $\num{-17.4}$  \\
    $\omega$CDM              & $\num{7.1}$      & $\num{7.4}$     &  $\num{14.9}$  &   $\num{1.5}$   & $\num{20.4}$    & $\num{7.1}$ \\
    $\omega_0 \omega_a$CDM   & $\num{18.4}$     & $\num{11.1}$ & $\num{47.0}$   &    $\num{1.5}$  & $\num{45.4}$   & $\num{8.1}$  \\
    \midrule
    \textbf{$\bm{\nu}$CKN with}\\
    $\Lambda$CDM             & $\num{-3.7}$     & $\num{-1.0}$    & $\num{-18.1}$  &   $\num{-9.4}$  & $\num{-43.2}$   & $\num{-19.0}$ \\
    $\omega$CDM              & $\num{5.1}$      & $\num{2.5}$    & $\num{13.3}$ &    $\num{1.5}$  & $\num{19.6}$  & $\num{5.5}$ \\
    $\omega_0 \omega_a$CDM   & $\num{16.4}$     & $\num{6.2}$  & $\num{45.4}$ &     $\num{1.5}$ & $\num{44.6}$   & $\num{6.5}$ \\
    \bottomrule
    \end{tabular}
    \label{tab:fp}
    \end{center}
\end{table}
\begin{table}
    \caption{Shown is the future projection of the results of the DESI experiment after five years of measurements (DESI-5Y) and the Euclid experiment. We convert the $\chi^2_\text{min}$ difference of the $\nu$CKN and $\Lambda$CDM model in table~\ref{tab:fp} into significances $\Sigma$.}
    \begin{center}
    \begin{tabular}{c c c c c c c} 
    \toprule
    \multirow{2}{*}{\bf Models} & \multicolumn{2}{c}{DESI-5Y} & \multicolumn{2}{c}{Euclid-Unc}& \multicolumn{2}{c}{DESI-5Y + Euclid-Unc}\\
    \cmidrule(l){2-3} 
    \cmidrule(l){4-5}
    \cmidrule(l){6-7}
    & $\Sigma_\text{DESY5}$ & $\Sigma_\text{Pantheon+}$ & $\Sigma_\text{DESY5}$  & $\Sigma_\text{Pantheon+}$ & $\Sigma_\text{DESY5}$  & $\Sigma_\text{Pantheon+}$\\
    \midrule
    \textbf{$\bm{\nu}$CKN with}\\
    $\Lambda$CDM             & $\num{-1.9}\sigma$     & $\num{-1.0}\sigma$    & $\num{-4.3}\sigma$  &   $\num{-3.1}\sigma$  & $\num{-6.6}\sigma$   & $\num{-4.4}\sigma$ \\
    \bottomrule
    \end{tabular}
    \label{tab:fp2}
    \end{center}
\end{table}

\section{Summary and Discussion}
\label{sec:dis}
In this study, we have explored the cosmological consequences of the CKN bound and performed a global analysis 
of cosmological data from the DESI BAO analysis combined with DES-SN5YR (DESY5) or Pantheon+ supernova data.
In order to account for unknown prefactors and to include alternative models 
where the dark energy density scales with $H^2$, we have generalized the CKN model to the $\nu$CKN model.
Due to the overlaps between the DESY5 and 
Pantheon+ datasets, we have considered only one supernova dataset at a time. 
Remarkably, the CKN model is found to provide a good fit to the data, 
even preferred over the $\Lambda$CDM model. Despite the additional model parameter, the $\nu$CKN
model did not provide a better fit compared to the CKN model. For the DESI BAO data combined with the 
Pantheon+ dataset, the CKN model provides a better fit 
than all alternative models considered in this work. 
These results demonstrate the need for a more comprehensive analysis, possibly incorporating 
power spectra and lensing information from CMB measurements and assessing compatibility with the early universe through big bang 
nucleosynthesis data. Moreover, new data expected to be released in the next few years will improve the statistics dramatically.
A projection of the change of the presently allowed parameter regions with the projected improved statistics justifies the expectation
that it will be possible to discriminate the ($\nu$)CKN model from the $\Lambda$CDM model with a statistical
significance up to $6.6\sigma$.

In summary, the CKN bound that addresses the restriction gravity poses on the validity of QFT has intriguing cosmological consequences.
These consequences are now starting to be probed by new cosmological data.

\section*{Acknowledgements}
P.A. was supported by the \textit{Cusanuswerk} during the early stages of this work, and later on by the \textit{Stu\-di\-en\-stif\-tung des deutschen Volkes}.
M.H. acknowledges support by grants
PID2020-113775GB-I00 (AEI/10.13039/ 501100011033) and CIPROM/2021/054
(Generalitat Valenciana).


\end{document}